\documentclass{aa} 

\usepackage{graphicx}  
\usepackage{txfonts}  
\usepackage{natbib}
\bibpunct{(}{)}{;}{a}{}{,}

\begin{document}  
 
\title{Fingering Convection in Red Giants Revisited}
  
  \author{ F. C. Wachlin\inst{1},
           S. Vauclair\inst{2,3},
           \and
           L. G. Althaus\inst{1}
           }
\institute{Instituto de Astrof\'{\i}sica de La Plata (UNLP - CONICET). 
           Facultad de Ciencias Astron\'omicas y Geof\'{\i}sicas. 
           Universidad Nacional de La Plata - Argentina\\
           \email{fcw@fcaglp.unlp.edu.ar}     
           \and
           Universit\'e de Toulouse, UPS-OMP, IRAP, France
           \and
           CNRS, IRAP, 14 avenue Edouard Belin, 31400 Toulouse, France\\
           }
\date{\today}

\abstract
{Fingering (thermohaline) convection has been invoked for several years 
as a possible extra-mixing which could occur in Red Giant stars due to 
the modification of the chemical composition induced by nuclear reactions 
in the hydrogen burning zone. Recent studies show however that this mixing 
is not sufficient to account for the needed surface abundances.}
{A new prescription for fingering convection, based on 3D numerical 
simulations has recently been proposed (BGS). The resulting mixing 
coefficient is larger than the ones previously given in the literature. 
We compute models using this new coefficient and compare them to previous
studies.}
{We use the LPCODE stellar evolution code with the GNA generalized version 
of the mixing length theory to compute Red Giant models and we introduce 
fingering convection using the BGS prescription.}
{The results show that, although the fingering zone now reaches the outer
dynamical convective zone, the efficiency of the mixing is not enough
to account for the observations. The fingering mixing coefficient should 
be increased by two orders of magnitude for the needed surface abundances
to be reached.}
{We confirm that fingering convection cannot be the mixing process needed 
to account for surface abundances in RGB stars.}

  
\titlerunning{Fingering Convection in Red Giants Revisited}
  
\authorrunning{Wachlin et al.}  

\maketitle 

\section{Introduction}  
\label{intro}

The formation and evolution of the chemical elements and 
their isotopic ratios in the Galaxy is a very rich and complex subject. 
At the present time, precise observations are obtained not only in stars 
with spectroscopic methods, but also in our close neighborhood, due to 
the detailed analyses of the pre-solar grains found inside meteorites 
and comets \citep{2007HiA....14..339A, 2007HiA....14..341O,
2007HiA....14..345L, 2007HiA....14..357N}.
The element abundances 
and isotopic ratios measured in these grains may give information 
on the sites where these elements were processed. It has been shown 
that they come from evolved stars, RGB and mostly AGB stars. 
The comparisons between the observed abundance ratios and the abundance 
variations computed in stellar models show evidences that extra-mixing 
must occur in these stars, besides standard convective zones 
\citep{2003ApJ...582.1036N}.

Evidence for the need of an extra-mixing process in RGB stars was 
already given by \citet{1995ApJ...453L..41C}. She pointed out that the 
observations of the carbon isotopic ratio in low mass Red Giant Stars 
suggest that a mixing mechanism occurs below the convective zone, 
after the first dredge-up, at the moment when the hydrogen burning shell 
(HBS) reaches the region which was mixed during the dredge-up. 
Such a mixing could also partially destroy $^3$He, which would account 
for the observations of this element in HII regions, less abundant 
than expected from its formation in main sequence stars.

Further computations by \citet{2006A&A...453..261P} showed that 
consistent computations of rotation-induced mixing in the framework 
of the shellular approximation \citep{1992A&A...265..115Z} could not 
lead to a strong enough mixing to account for the observations. 
\citet{2006Sci...314.1580E} proposed that the mean molecular weight 
decrease induced by nuclear reactions in the hydrogen burning
shell could lead to hydrodynamical instabilities and help account for the
observations. They assumed however that the effect was dominated by
Rayleigh-Taylor mixing, which is not the case. \citet{2007A&A...476L..29C}
pointed out that thermohaline convection was the first process occurring in the
presence of inverse $\mu$-gradients. They computed this effect using the
prescription proposed by \citet{1972ApJ...172..165U} for the mixing coefficient
and found that the abundance observations could nicely be reproduced.
Unfortunately, it was later proved by numerical simulations and analytical
computations that the mixing coefficient they used was strongly overestimated
\citep{2010ApJ...723..563D, 2011ApJ...728L..29T, 2011A&A...533A.139W,
2012ApJ...753...49V}.
Later on, an attempt was made to treat thermohaline and rotational-induced mixing
together, by adding the corresponding mixing coefficients
\citep{2010A&A...522A..10C,2011A&A...536A..28L}. However this very simple
 treatment did not take into account the influence of horizontal turbulence
which would reduce the effect of the thermohaline process, as mentioned by
\citet{2012ApJ...753...49V}. This was later recognized by
\citet{2013A&A...553A...1M}.

At the present time, we are in a situation where the presence of an 
extra mixing in Red Giants is clearly needed but the reason for this 
mixing is still unknown \citep{2003ApJ...593..509D, 2007ApJ...671..802B,
2009ApJ...696.1823D, 2010MNRAS.403..505S, 2011ApJ...727L...8D}.
We also know that specific hydrodynamical processes in AGB stars 
are needed to explain the chemical evolution of the Galaxy 
\citep{2003MNRAS.340..722D, 2007A&A...464L..57S, 2008ApJ...677..556T, 
2009MNRAS.396.2313S}.

In this paper, we focus on the subject of thermohaline mixing in RGB stars.
\citet{2013ApJ...768...34B} performed new 3D-simulations
and gave a new 1D-prescription of thermohaline convection (rather referred to
now as ``fingering convection'' as ``thermohaline'' is more appropriate
for the ocean than for stars). They gave evidence that 
\citet{2011ApJ...728L..29T} treatment underestimated the mixing efficiency
in the limits of very small Prandtl and Lewis numbers, which are 
characteristic of stellar conditions. It seemed important to test this 
new prescription for the Red Giant case. This was the motivation of the
present paper.
We present numerical computations of thermohaline
convection using this new coefficient (Sect. \ref{simulations}). 
The results are given in Sect. \ref{results}. We show that, although 
the mixing is now clearly more efficient than obtained with previous 
coefficients, it is not sufficient to lead to the needed
abundance dilution of $^3$He. These results and their implications 
are discussed in Sect. \ref{discussion}.


\section{Numerical computations}  
\label{simulations}
\subsection{Stellar models}

We computed  the evolution of a $0.9 M_\odot$  model with initial
metallicity  of [Fe/H]$=-1.3$ from  the zero-age
main sequence (ZAMS) until the upper red giant branch (RGB), where the
luminosity is $L\approx 10^3 L_\odot$. 
We also computed stellar models with different metallicities, 
one with [Fe/H]$=-0.3$ (almost solar) and one with 
[Fe/H]$=-2.3$ to check the 
influence of the chemical composition on our final results. 
All calculations have been done using the LPCODE stellar evolutionary
code \citep{2005A&A...435..631A, 2013A&A...557A..19A}. This is a 
well-tested and  calibrated code that  has been amply used  in the
study  of different  aspects of  low-mass star  evolution, including
white  dwarf stars.  In particular,  and  for the  relevance of  the
present  paper, the  code includes  a  generalized version  of the
mixing length theory developed by \citet[GNA
hereafter]{1993ApJ...407..284G} \citep[see also][]{1996MNRAS.283.1165G}.
This double  diffusive  convection
theory has already been successfully  implemented in a similar context
by \citet{2011A&A...533A.139W}, and we  refer the reader  to this
paper for details about the implementation  of the GNA theory in our
code. We set the GNA mixing length parameter to an equivalent value
of $\alpha=1.66$ in the classical (MLT) theory. 
The use of the GNA convection theory allows us to infer the different 
unstable transport regimes, namely, dynamical convection, 
semi-convection, and fingering (thermohaline) convection, 
and to treat the corresponding mixing processes by using implemented 
diffusion coefficients. 

\begin{figure}
\resizebox{\hsize}{!}{\includegraphics{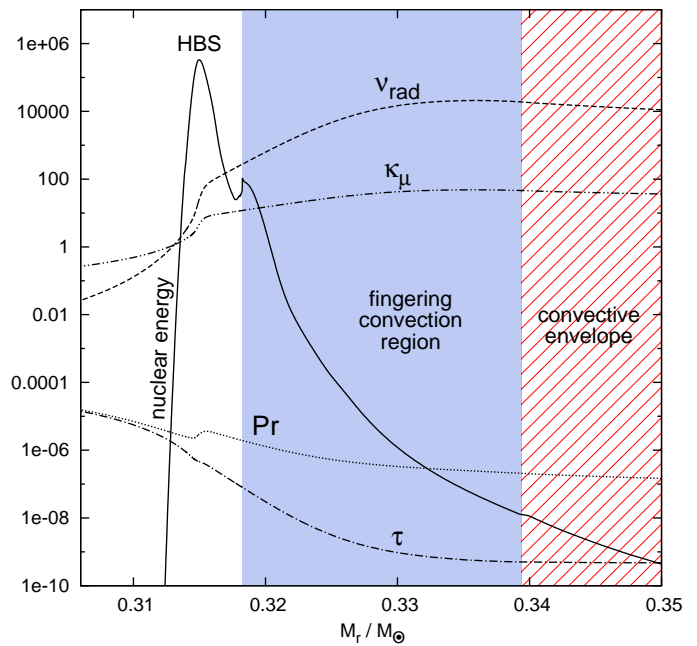}}
\caption{Profiles of some relevant parameters for fingering convection 
as a function of the internal mass in a $0.9 M_\odot$  model 
with [Fe/H]$=-1.3$. The location of the HBS is also shown 
(nuclear energy is in erg g$^{-1}$ s$^{-1}$). 
See text for details.}
\label{fig:kappamu-etc}
\end{figure}

Figure \ref{fig:kappamu-etc} displays the variations with depth 
of several parameters which are important for the computation 
of fingering convection, 
in a $0.9 M_\odot$  model with [Fe/H]$=-1.3$. These parameters are the 
radiative viscosity $\nu_{rad}$, the molecular diffusivity $\kappa_{\mu}$, 
the Prandtl and inverse Lewis numbers. The nuclear energy production is 
also presented in the same graph, to materialize the hydrogen burning 
zone. This model corresponds to the moment when the fingering 
region reaches the bottom of the envelope dynamical convective zone 
(see below, Sect. \ref{behaviorFCZ}). If we compare this graph 
with the values given
by \citet{2010ApJ...723..563D} in his Table 2, we see that they agree 
only in the lower part of the HBS. In this region the
radiative viscosity $\nu_{rad}$ is of the same order as the molecular
diffusivity $\kappa_{\mu}$, but their values are very different above
and below these layers. At the place where the fingering 
convection actually develops, the radiative viscosity is more than two 
orders of magnitude larger than the molecular diffusivity, which 
strongly modifies the Prandtl number. 

We also computed stellar models in which we included overshooting, following
the prescription of \citet{1996A&A...313..497F}. 
All convective 
boundaries are extended by assigning
to that region an exponentially decaying diffusive coefficient
of the form
\begin{equation}
 D=D_0 \exp\left( -\frac{2\, z}{f\, H_\mathrm{p}} \right),
\end{equation}
where $D_0$ is the diffusive coefficient near the edge of the 
convective zone, $z$ denotes the distance of the considered layer to this edge,
$H_\mathrm{p}$ is the pressure scale 
height and $f$ is a measure of the efficiency of the extra 
partial mixing. In the following, we explore three cases of overshooting: 
moderate ($f=0.015$), intermediate ($f=0.075$) and extreme 
($f=0.15$).

\subsection{Treatment of fingering convection}

We use in this paper the recent prescription   given  by
\citet{2013ApJ...768...34B} (hereafter BGS) for the computation of fingering
convection, which represents a real improvement compared to the previous
treatments \citep[see also][]{2014arXiv1407.1437Z}.

Fingering (thermohaline) convection is a well-known process in the ocean. This
instability occurs when hot salted water comes upon cool fresh water. It is
indeed at the origin of the global circulation in the Earth Ocean, called
"thermohaline circulation". In stars, a similar instability occurs every time
heavy matter comes upon lighter one, in the presence of a stable temperature
gradient. This may happen in the case of accretion of planetary matter
\citep{2004ApJ...605..874V, 2011ApJ...728L..30G, 2013A&A...557L..12D}, of
accretion of matter from a companion
\citep{2008MNRAS.389.1828S, 2008ApJ...677..556T}, in the case of a local
$\mu$-decrease due to nuclear reactions as in Red Giants
\citep{2007A&A...476L..29C}, or in the presence of iron-rich layers
induced by atomic diffusion \citep{2009ApJ...704.1262T,
2014arXiv1407.1437Z}.

The first treatments of fingering convection in stars were purely analytical
\citep{1972ApJ...172..165U,1980A&A....91..175K}. They differed by orders of
magnitude, according to the assumed shape of the "fingers", which was unknown.
Contrary to \citet{2004ApJ...605..874V}, \citet{2007A&A...476L..29C}
used the \citet{1972ApJ...172..165U} value, much larger than the 
\citet{1980A&A....91..175K} one. More
recently, 2D and 3D numerical simulations were performed, all converging on the
result that the \citet{1972ApJ...172..165U} value was strongly overestimated
\citep{2010ApJ...723..563D,2011ApJ...728L..29T}. The new simulations by BGS
including the evolution of the fingers with time, and the associated
prescription, give coefficients slightly larger than the previous ones. This is
what is used in the present computations.

\begin{figure}
\resizebox{\hsize}{!}{\includegraphics{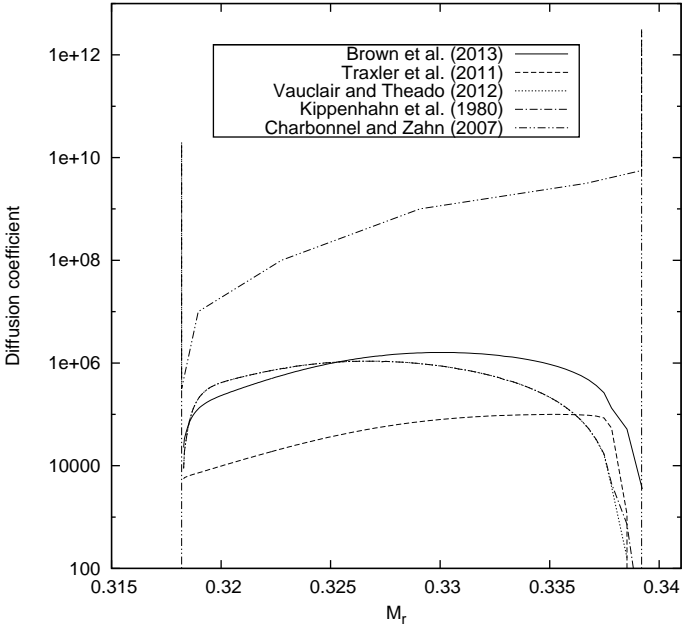}}
\caption{Comparison of the diffusion coefficients for the 
fingering region corresponding to different prescriptions
used in the past. The coefficients correspond to a stellar 
model at the time the fingering zone reaches the bottom 
of the dynamical convective zone
when the BGS prescription is used
(right panel of Fig. \ref{fig:beforeafterbump}.)}
\label{fig:diffusion-coefs}
\end{figure}

Figure \ref{fig:diffusion-coefs} displays the coefficients 
obtained for various prescriptions in a 0.9 $M_\odot$ model 
with [Fe/H]$=-1.3$. As will be seen below, the use of the BGS 
coefficient is enough for the fingering region to
reach the bottom of the CZ, but not enough to reduce
$^3$He significantly. Figure \ref{fig:diffusion-coefs} corresponds to 
the time at which the fingering instability reaches the base
of the convection zone when the BGS coefficient is used. 

\section{Results}  
\label{results}

\subsection{Behavior of the fingering convection zone}
\label{behaviorFCZ}  

We first describe the results obtained for a 0.9 $M_\odot$ model 
with [Fe/H]$=-1.3$, without overshooting. We are mainly interested in 
the RGB phase around the luminosity bump, when the advance of the 
HBS over the homogeneous region left by the first dredge-up triggers 
fingering instability. This double-diffusive instability is closely 
related to the appearance of a compositional gradient inversion 
($\nabla_\mu<0$) soon after the luminosity bump.

\begin{figure}
\resizebox{\hsize}{!}{\includegraphics{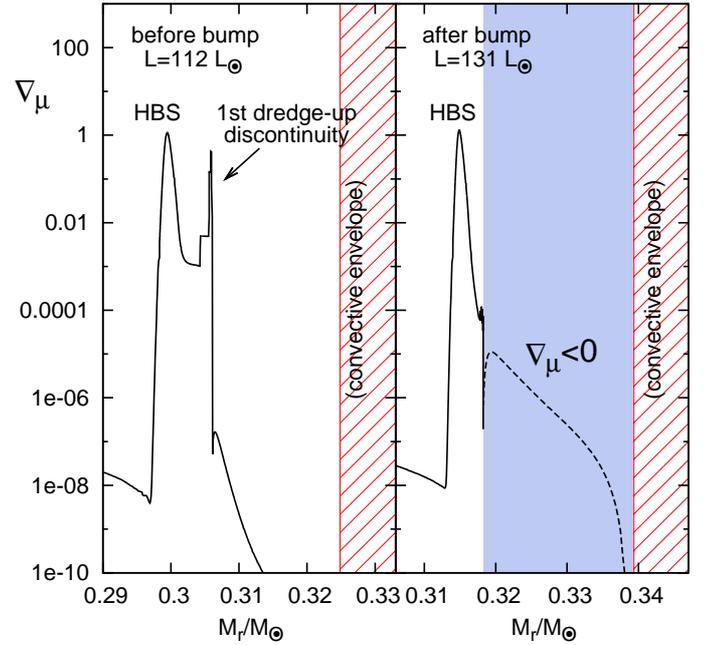}}
\caption{Profile of the mean molecular weight gradient $\nabla_\mu$
as a function of mass. Solid line stands for $\nabla_\mu>0$ regions
and the dashed line otherwise. Left and right panels correspond
to the situation before and after the bump, respectively, for 
the model of $0.9 M_\odot$, [Fe/H]$=-1.3$. HBS: 
hydrogen burning shell; shaded zone: fingering unstable region;
line hatch pattern: innermost part of the convective envelope. The left
and right panels are separated by 9.7 Myr of evolution.}
\label{fig:beforeafterbump}
\end{figure}

Figure \ref{fig:beforeafterbump} shows the compositional
gradient profile near the location of the HBS for two evolutionary
stages before (left panel) and after (right panel) the luminosity bump. 
In the left panel two peaks are apparent, the deeper peak 
(at $M_r\approx 0.3 M_\odot$) corresponds to the region 
where H is depleted by nuclear reactions in the HBS, whereas the second
peak (at $M_r\approx 0.306 M_\odot$) corresponds to the 
chemical discontinuity at the point of maximum penetration 
of the first dredge-up. Since the HBS moves to the surface as 
H becomes exhausted, the situation corresponds to
a moment shortly before the occurrence of the luminosity bump. 
No compositional gradient 
inversion is evident. In the right panel the situation is quite different.
The panel  
illustrates the moment when the fingering instability region (shaded zone) 
first touches the convective envelope, 9.7 Myr after the left panel 
situation. It can be seen that the HBS has now reached the 
former homogeneous region. A compositional
gradient inversion (dashed line) has developed between the HBS and 
the convective envelope. This $\nabla_\mu<0$ zone starts soon after 
the luminosity bump as a small region at the external
wing of the HBS, but grows up progressively until it reaches the receding
convective envelope.  

The most remarkable result here is that the fingering unstable 
zone eventually reaches the
outer convection zone, which is a clear result of using the new prescription
given  of \citet{2013ApJ...768...34B} for the fingering convection.
It never happened with previous mixing coefficients, except the one used 
by  \citet{2007A&A...476L..29C}, which was strongly overestimated.
This is the most important prediction of our simulations. 

However, although the contact between both unstable regions might 
provide the
extra-mixing mechanism that allows convective envelope material to
reach the HBS and thus modify the surface abundances of $^3$He, $^7$Li,
$^{12}$C, $^{13}$C and $^{14}$N, no change of these abundances 
was found in this simulation. The reason is related to the efficiency of the
thermohaline mixing, which decreases according to the decrease of the
$\mu$-gradient, so that when the mixed zone reaches the classical 
convective zone the fingering mixing efficiency becomes 
too weak to modify the abundances (see Figs. 
\ref{fig:abund-Li-etc} and \ref{fig:abund-He3}, 
full line labelled BGS).

\subsection{Influence of overshooting and different metallicities}

We have checked the possible influence of overshooting on the 
preceding results. Extending the envelope convective zone could indeed 
lead to an overall more efficient mixing and help changing the 
surface abundances as needed. We performed three new experiments, 
introducing a moderate ($f=0.015$), an intermediate ($f=0.075$) and 
a strong ($f=0.15$) overshooting below the dynamical convective zone. 
No changes in the surface abundances were obtained in any of these 
three cases. The mixing efficiency of the fingering convection
is much too low for the elements to be mixed between the 
convective envelope and the HBS, even in the presence of overshooting. 

We also computed models with two different metallicities, 
[Fe/H]$=-0.3$ and  [Fe/H]$=-2.3$. 
In both experiments the behavior of the 
fingering zone is very similar to that of the [Fe/H]$=-1.3$ case, 
although some small differences do appear. In the less metallic 
model the fingering region grows with a timescale of about 
10 Myr but never reaches the bottom of the convective zone, 
whereas for the more metallic model the fingering region 
comes in contact with the convective zone but, 
in this case, the surface abundances shows a 
negligible change  (for example, $^{12}$C/$^{13}$C 
changes from 45.90 to 45.89). 

\subsection{Influence of artificially increasing the mixing coefficient}  

As a toy model, we decided to explore the consequences of
increasing the values proposed by \citet{2013ApJ...768...34B}
in order to check the impact of such a modification on the observed 
surface abundances.

\begin{figure}
\resizebox{\hsize}{!}{\includegraphics{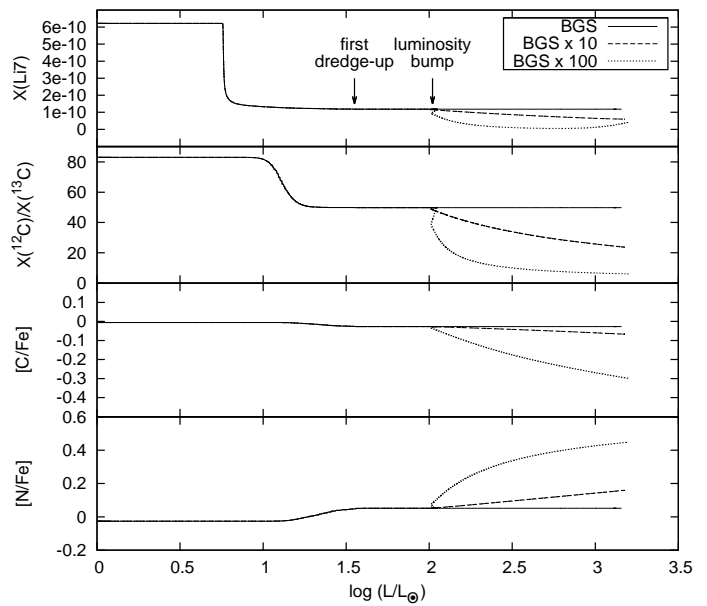}}
\caption{Evolution of the surface lithium abundance, of the carbon isotopic
ratio $^{12}$C/$^{13}$C, of [C/Fe] and [N/Fe] as a function of
the luminosity logarithm for three simulations where the 
diffusion coefficient has been changed. The
full line (labeled ``BGS'') was obtained implementing 
the \citet{2013ApJ...768...34B} prescription, the dashed line
(labeled ``BGS x 10'') corresponds to the case where the 
diffusion coefficient is artificially increased by a factor of 10 and
the dotted line (labeled ``BGS x 100'') refers to computation 
with a diffusion coefficient 100 times larger than that of 
\citet{2013ApJ...768...34B}. The luminosity of the model at the
moment of the first dredge-up and at the start of the luminosity bump
are marked at the top figure.
}
\label{fig:abund-Li-etc}
\end{figure}  

We artificially increased
\citet{2013ApJ...768...34B} diffusion coefficients by factors 
of 10 and 100, respectively. No overshooting was introduced in 
these computations. Figure \ref{fig:abund-Li-etc} shows the results, 
including the outcome of the simulation obtained with the original 
BGS diffusion coefficients. 

An increase of the mixing efficiency of the fingering region by 
a factor of 10 is enough to modify the surface abundances of some 
elements, particularly the carbon isotopic ratio ($^{12}$C/$^{13}$C). 
Other quantities like the mass fraction of $^7$Li, or 
[C/Fe] and [N/Fe] present a small change at the luminosity bump 
($\log L_\mathrm{bump}/L_\odot\approx 2.05$), and in the final abundances. 

For a diffusion coefficient multiplied by 100, the surface abundances
change more rapidly than in the former case. The final values are 
clearly different from the ones before the luminosity bump. 
The modification of the carbon isotopic ratio is particularly abrupt, 
compared with the evolution of the other indicators.

\begin{figure}
\resizebox{\hsize}{!}{\includegraphics{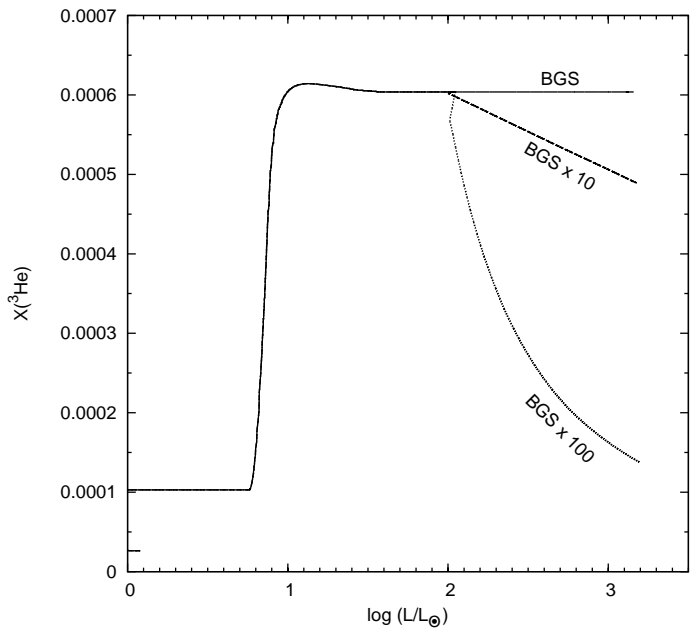}}
\caption{Evolution of the surface abundance of the mass fraction
of $^3$He. The labels of the curves are as in 
Fig.\ref{fig:abund-Li-etc}.}
\label{fig:abund-He3}
\end{figure}

Of particular interest is the evolution of the surface $^3$He 
abundance for the different numerical experiments mentioned 
before. While no changes appear when we use BGS prescription, 
the surface abundances are modified when we introduce 
larger mixing rates. Figure \ref{fig:abund-He3} shows the 
evolution of the $^3$He surface abundance for the same 
simulations as presented in Fig. \ref{fig:abund-Li-etc}. In the
classical picture, the surface $^3$He composition changes 
as a consequence of the first dredge-up. After that episode,
in the absence of any other mixing process, it 
should remain constant for the rest of the RGB evolution. Figure
\ref{fig:abund-He3} shows that this is even the case for the
implementation of BGS prescription. However, as the mixing 
coefficient is increased, the surface $^3$He mass fraction 
decreases as a consequence of its consumption
in the HBS. For a diffusion coefficient multiplied by 10, we see
that X($^3$He) decreases by a small amount, from $0.604\times 10^{-3}$ 
to $0.490\times 10^{-3}$, whereas for an increase by a factor 100, 
the $^3$He mass fraction rapidly decays to a final value of 
$0.137\times 10^{-3}$, close to its initial abundance and needed
to reconcile the $^3$He abundance with that observed in HII 
regions.

\section{Discussion and conclusion}
\label{discussion}

Since the \citet{2007A&A...476L..29C} paper in which fingering 
(thermohaline) convection was invoked as a possible explanation 
of the surface abundances in Red Giant stars, the description 
of this instability has much improved owing to 3D-numerical 
simulations. It has been shown by several authors 
\citep{2010ApJ...723..563D,2011ApJ...728L..29T,2012ApJ...753...49V}. 
that the Ulrich prescription used at the beginning was strongly 
overestimated. All recent studies \citep[see][]{2011A&A...533A.139W}
converge on the result that this extra-mixing is not the right 
process able to account for the observations. 

The present paper was motivated by the most recent 3D-simulations 
of fingering convection and the derived 1D-prescription of BGS. 
These new simulations and the new prescription give a better 
treatment than the previous ones for stellar conditions. The 
resulting mixing coefficient is larger and it was interesting 
to test its influence on the general results. 

We have computed Red Giant models with several metallicities, 
with or without overshooting below the envelope dynamical 
convective zone. The important new result compared to 
previous studies is that, with this new prescription, the 
fingering zone induced by nuclear compositional changes 
may reach the bottom of the dynamical convective zone, 
which was not the case before. However, the efficiency of the 
added mixing induced by fingering convection is still two 
orders of magnitude too low to account for the observations. 

We confirm in this paper that another kind of extra mixing 
is needed to account for the chemical composition of Red Giant 
and reconcile the production of $^3$He with the galactic observations.  

\begin{acknowledgements}
This  research   was  supported, in part   by  PIP
112-200801-00940 from  CONICET and by AGENCIA through the Programa de
Modernizaci\'on Tecnol\'ogica BID 1728/OC-AR. FCW and LGA would 
like to thank Marcelo M. Miller Bertolami for useful 
discussions about RGB stars and for major improvements made to LPCODE.
\end{acknowledgements}  
  
\bibliographystyle{aa} 
\bibliography{rgb2014} 

\end{document}